\documentclass[sn-mathphys,Numbered]{sn-jnl}

\usepackage{graphicx}%
\usepackage{multirow}%
\usepackage{amsmath,amssymb,amsfonts}%
\usepackage{amsthm}%
\usepackage{mathrsfs}%
\usepackage[title]{appendix}%
\usepackage{xcolor}%
\usepackage{textcomp}%
\usepackage{manyfoot}%
\usepackage{booktabs}%
\usepackage{algorithm}%
\usepackage{algorithmicx}%
\usepackage{algpseudocode}%
\usepackage{listings}%

\newcommand{\ket}[1]{\left|#1\right\rangle}
\newcommand{\bra}[1]{\left\langle #1\right|}

\newcommand\tab[1][1cm]{\hspace*{#1}}

\theoremstyle{thmstyleone}%
\theoremstyle{thmstyletwo}%

\theoremstyle{thmstylethree}%

\begin{document}


\title[High-dimensional graphs convolution  for quantum walks   photonic applications]{High-dimensional graphs convolution  for quantum walks photonic  applications}

\author*[1]{\fnm{Roman} \sur{Abramov}}\email{rk.roman.abramov@gmail.com
}
\affil[1]{\orgname{Technical University of Munich}, \orgaddress{\street{Arcisstraße 21}, \city{Munich}, 
\postcode{80333}, \country{Germany}}}
\equalcont{These authors contributed equally to this work.}


\author[2]{\fnm{Leonid} \sur{Fedichkin}}\email{leonidf@gmail.com}
\affil[2]{\orgdiv{Valiev Institute of Physics and Technology}, \orgname{Russian Academy of Sciences}, \orgaddress{\street{Street}, \city{Moscow}, \postcode{117218}, \country{Russia}}}
\equalcont{These authors contributed equally to this work.}

\author[3,4]{\fnm{Dmitry} \sur{Tsarev}}\email{dmitriy\_93@mail.ru}

\affil[3]{\orgdiv{Institute of Advanced Data Transfer}, \orgname{ITMO University}, \orgaddress{\street{Kronverksky Pr. 49, bldg. A,}, \city{St. Petersburg}, \postcode{197101}, \country{Russia}}}

\affil[4]{\orgdiv{Quantum Light Engineering Laboratory, Institute of Natural and Exact Sciences}, \orgname{South Ural State
University}, \orgaddress{\street{Lenin Av., 76}, \city{Chelyabinsk}, \postcode{454080}, \country{Russia}}}

\author*[3]{\fnm{Alexander} \sur{Alodjants}}\email{alexander\_ap@list.com}

\equalcont{These authors contributed equally to this work.}


\abstract{
Quantum random walks represent a powerful tool for the implementation of various quantum algorithms. 
We consider a convolution problem for the graphs which provide quantum and classical random walks. We suggest a new method for lattices and hypercycle convolution that preserves quantum walk dynamics. Our method is based on the fact that some graphs represent a result of Kronecker's product of line graphs. We support our methods by means of various numerical experiments that check quantum and classical random walks on hypercycles and their convolutions.  
Our findings may be useful for saving a significant number of qubits required for algorithms that use quantum walk simulation on quantum devices.}




\maketitle

\section{Introduction}\label{sec1}

Random walks represent a fruitful tool for modeling probabilistic processes in physics \cite{Kac} and beyond \cite{Kutner, bartumeus}.  In information science random walks on graphs possess various applications in algorithms, like PageRank, etc. \cite{Xia}.  Quantum walks (QWs), establishing quantum generalization of classical random walks, are potentially quadratically faster due to quantum interference and path entanglement phenomena \cite{Kempe}. These features make QWs useful for quantum computing purposes \cite{Childs, Venegas, Ambainis, Anadu}.  QWs are at the heart of  Grover's quantum search algorithm \cite{Portugal}, explored for a speedup problem in various algorithms of quantum machine learning  \cite{Biamonte, Melnikov2023}.

It is noteworthy, that the advantages of  QW are proven for simple graphs, which may be established by line \cite{Ambainis2001, Aharonov}, cycle \cite{Solenov}, hypercube \cite{Krovi, Kempe2005, Makmal}, complete and glued trees graphs \cite{Santos, Childs2003}. However, the speedup problem is still unsolved for arbitrary graphs.  This problem is significant for the design of quantum computers capable of quantum advantages in the NISQ era, e.g. \cite{Preskill}. 
Obviously, such a problem can be solved in various ways.

In our works \cite{Melnikov2019, Melnikov2020}, we showed that the speedup problem for arbitrary graphs depends on the graph peculiarities, detection procedure, and decoherence effects. In particular, we designed a classical-quantum convolutional neural network (CQCNN) that enables us to recognize whether the classical or quantum walk is faster for a given adjacency matrix that determines an unknown graph.  

In this work, we suggest another approach to solve the problem of quantum walks advantage in the NISQ era. It is based on the reduction (convolution) of the original graph for a smaller one, for which the quantum walk speedup problem can be easily estimated, or is generally known.

The importance of this problem to the design of quantum computing hardware is clear. The QWs mapped as quantum circuits require $2^n$ vertices to represent $n$ qubits,  e.g.   \cite{Douglas, Portugal2022}. 
At the same time, mapping the target Hamiltonian to the quantum hardware graph (minor embedding procedure) represents an important prerequisite for current quantum computation devices, e.g. \cite{Choi}. Such a procedure is necessary due to dimension constraints that undergo real-world physical systems proposed for quantum computing. Thus,
for efficient simulation of large graphs, it is necessary to implement the reduction of the number of qubits and depth of quantum circuits as much as possible. 

Photonic circuits represent one of the promising tools for quantum computing now \cite{Peruzzo}.
Low-loss optical waveguides effectively possess only one \cite{Silberberg, Szameit}, or two dimensions \cite{Tang} for photon quantum walks purposes.  The third dimension is assigned for the propagation coordinate, which is associated with the walking time, or algorithm implementation physical time.    

However, not all graphs admit simple and unique reduction, if it exists at all. 
One possible way to solve a given problem is to reduce an unknown graph to a known geometric 
construction, for which random walks are well known. 
In particular, in \cite{maczewsky2020} authors demonstrated how high-dimensional networks in some cases might be mapped on 1D waveguide chains possessing various coupling coefficients between the waveguides, e.g. \cite{yu2016}. 

In this work, we consider single-particle continuous-time quantum walks (CTQW) performed on various graphs, which pose evident geometric interpretations and may be mapped onto the waveguide arrays that represent photonic quantum computation hardware. In this regard we restrict ourselves by the Markovian approach to QWs characterization, e.g. \cite{Melnikov2020} and cf. \cite{Kutner}.
In Sec. 2 we examine graph convolution problems for various geometric constructions, starting from the well-known line, hypercube, and circle graphs. Then, in Sec. 3 we analyse more complicated and practically promising systems such as hypercycles and toruses. Our great interest in such systems relates to the possibility of using toric codes in fault-tolerant quantum computing, e.g.  \cite{Kitaev,Pedrocchi}.  
In Sec. 4  we establish numerical experiments of QWs simulations, which confirm our approach.
 Finally, in Sec. 5 we find out some specific peculiarities of the adjacency matrices that enable to mapping of considered structures onto the line graphs and lattices.  
The conclusion summarizes the results obtained.

\section{Quantum walks on hypercubes}

\subsection{General description}

Let us propose a quantum particle that is located at one of  $m$ positions on a graph with $m$ vertices, or stays in a superposition of these positions. We consider the quantum state of the particle in the form of an $m$-level system, that is  $\ket{\psi_D(t)} = \sum_{i=0}^{m-1} \alpha_i(t)\ket{i}$, where  $\lvert\alpha_i(t)\rvert^2$ is the probability of detecting the particle in vertex $i$  at time $t$. The evolution of this quantum state is governed by the Hamiltonian with nearest-neighbor hopping terms

\begin{equation}\label{Ham}
    \mathcal{H} = \hbar\Omega\sum_{i,j=0}^{m-1} A_{ij}\ket{i}\bra{j} = \hbar\Omega A, 
\end{equation}
where $A$ is an adjacency matrix of a graph that admits QW,  $A_{ij}$ are its matrix elements, and $\Omega$ is the hopping frequency. Below we set $\hbar=1$ for brevity.

The unitary evolution  of the quantum state represents a solution of the Schr\"{o}dinger equation, which is given by
\begin{equation}
    \ket{\psi_D(t)} = \mathrm{e}^{-i\Omega t A}\ket{\psi_D(0)}.
    \label{general_evolution}
\end{equation}
Notice, $A$ is not necessarily symmetric; the weights $A_{ij}$ are complex parameters and can, in general, lead to chiral QWs on weighted graphs~\cite{lu2016chiral}.

The Hamiltonian \eqref{Ham} experimentally can be easily realized for QWs performed in effectively 1D quantum optical systems \cite{Silberberg, Alodjants}.  
In this case one can speak about the realization of continuous-time quantum walks that occur due to photon propagation in low-loss ($0.1 dB/cm$ and below) tunnel-coupled waveguide arrays, see Fig.~\ref{fig:hypercube3}, and cf. \cite{Szameit}. The propagation distance along the waveguides plays the role of the time variable, see the right sketch in  Fig.~\ref{fig:hypercube3}. Parameter $\Omega$ characterizes maximal photon tunneling rate for optically interacting modes of nearest-neighbor waveguides, \cite{Kulik, Chen}.  Formally, we set below $\Omega=1$.  
This assumption corresponds to the normalization of the physical parameters and variables on $\Omega$. In particular,  the new time variable  $t$ that we use below corresponds to $\Omega t$, represented in \eqref{general_evolution}.  

However, the experimental realization of Hamiltonian \eqref{Ham} for high-dimensional graphs (arbitrary adjacency matrix $A_{ij}$) is cumbersome. Thus, the convolution of high-dimension graphs to a 1D physical system, which we consider below,  represents an important theoretical problem that possesses high practical impact and establishes our tasks of studies in photonics in the future.  

\subsection{Hypercubes convolution}

\begin{figure}[ht!]
	\centering
	\includegraphics[width=0.6\linewidth]{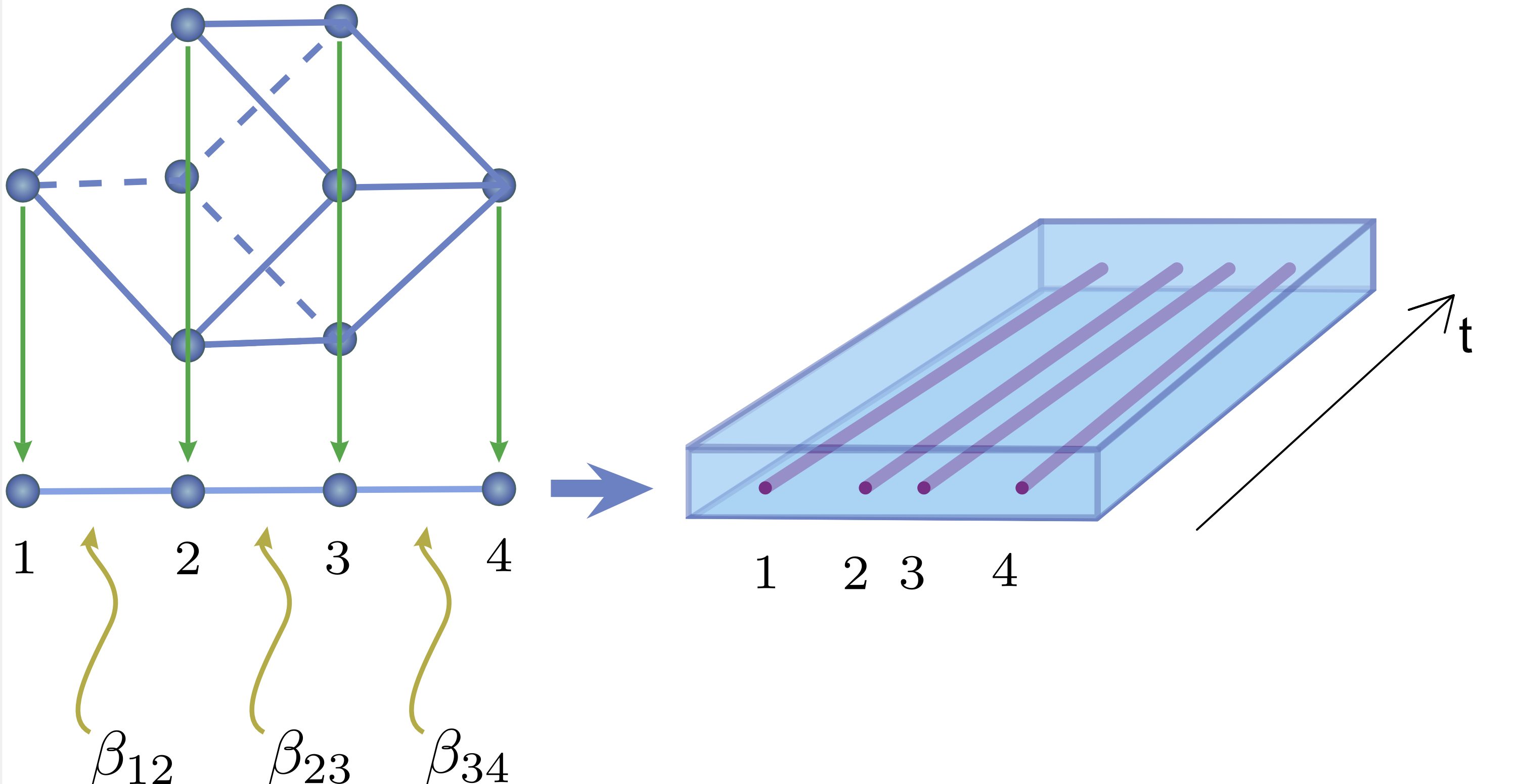}
	\caption{Mapping of cube graph onto a weighted line graph (left panel), and then, onto an array of optical waveguides (right panel). Dimensionless coupling coefficients between waveguides are $\beta_{12}=\beta_{34}=\sqrt{3}$, $\beta_{23}=2$, and may be tailored by using the distance between the waveguides.}
	\label{fig:hypercube3}
\end{figure}

\begin{figure}[ht!]
	\centering
	\includegraphics[width=0.6\linewidth]{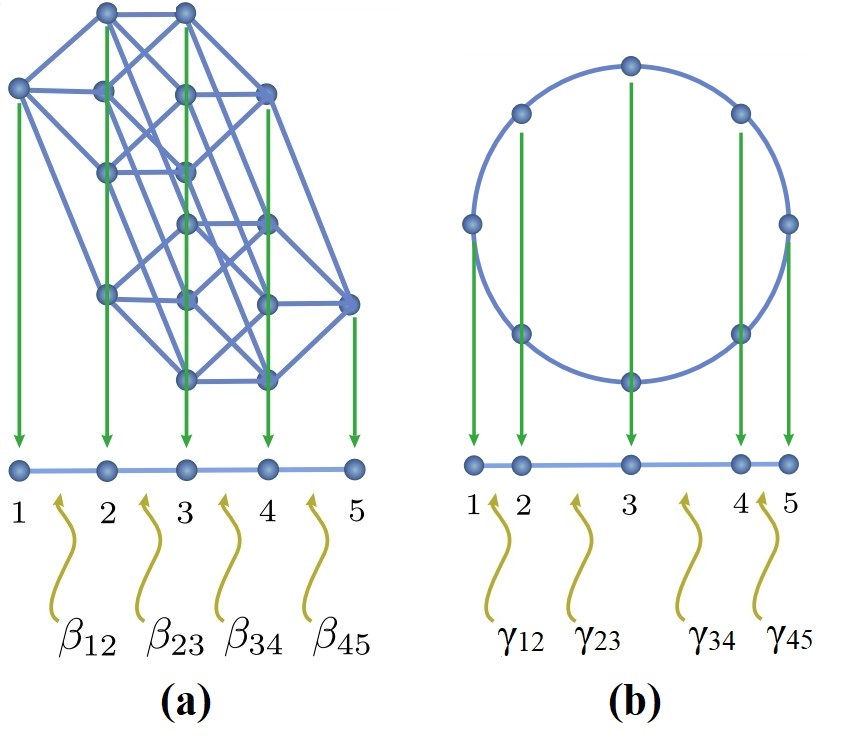}
	\caption{ Mapping of (a)  $D=4$ hypercube  and (b)  $D=1$ hypercycle (with $k=8$ vertices) onto the  weighted line graphs, respectively.}
	\label{fig:hypercube4}
\end{figure}

In this work, we exploit the fact that there is an equal probability of detecting a particle in vertices with the same Hamming distance for chosen graph structures. By taking this peculiarity into account, we can obtain a weighted line graph as a result of the hypercube mapping procedure for $D=3$   (Fig.~\ref{fig:hypercube3})  and for $D=4$ (Fig.~\ref{fig:hypercube4}), respectively. The Hamiltonian that governs QW in the mapped space can be defined for an arbitrary hypercube dimension $D$ as

\begin{equation}
    \mathcal{H}^\mathrm{hc\rightarrow line} = \sum_{i=1}^{D}\beta_{i,i+1}\big(\ket{i+1}\bra{i}+\ket{i}\bra{i+1}\big),
    \label{hypercube_to_line}
\end{equation}
where $\beta_{i,i+1}=\sqrt{i(D+1-i)}$ is a dimensionless  coupling coefficient for  vertices $i$ and $i+1$. It is noteworthy, that
there are only $(D+1)$ vertices, which are necessary for $D$-dimensional hypercube CTQW implementation. 

It is instructive to establish \eqref{hypercube_to_line} for some lower dimensions of $D$.
In the case of  hypercube with  $D=2$ (square)  \eqref{hypercube_to_line} implies
\begin{equation}\label{d2}
    \mathcal{H}_2^\mathrm{hc\rightarrow line} = \sqrt{2}\big(\ket{2}\bra{1} + \ket{1}\bra{2} + \ket{3}\bra{2} + \ket{2}\bra{3}\big).
\end{equation}


For a hypercube with  $D=3$ one can obtain (cf. Fig.~\ref{fig:hypercube3}),
\begin{equation}\label{d3}
\mathcal{H}_3^\mathrm{hc\rightarrow line} = \sqrt{3}\big(\ket{2}\bra{1} + \ket{4}\bra{3} + H.C.\big) \\ + 2\big(\ket{3}\bra{2} + \ket{2}\bra{3}\big),  
\end{equation}
where $H.C.$ denotes the Hermitian conjugated part.

Finally, in the  case of  hypercube for $D=4$, represented in Fig.~\ref{fig:hypercube4}, we get 
\begin{equation}\label{d4}
\mathcal{H}_4^\mathrm{hc\rightarrow line} = 2\big(\ket{2}\bra{1} + \ket{5}\bra{4}+ H.C.\big) + \sqrt{6}\big(\ket{3}\bra{2} + (\ket{4}\bra{3}+ H.C.\big). 
\end{equation}

Apart from the 2D case \eqref{d2} hypercube mapping in three and four dimensions \eqref{d3}, \eqref{d4} requires non-equal coefficients for the next neighbor vertices on the line graph. Practically, it may be realized by means of variation of the distance between the waveguides, e.g. \cite{Szameit}.

\section{Quantum walks on hypercycles}

\tab Hypercycles are high-dimensional cycles and are defined by their dimension $D$, number of vertices $k$ in each one-dimensional projection. Here we consider $k$ to be even.

\subsection{Hypercycle of $D=1$ convolution}

The Hamiltonian that governs the QW in the mapped space can be defined as:
\begin{equation} \label{hypercycle_d1}
    \mathcal{H}^\mathrm{hc\rightarrow line}= \sum_{i=1}^{\kappa}\gamma_{i,i+1}\big(\ket{i+1}\bra{i}+\ket{i}\bra{i+1}\big).
\end{equation}
Notice this is the same form of the Hamiltonian as in the case of hypercubes. The difference occurs for the values of couplings $\gamma_{ij}$ and in $\kappa$  that defines the number of terms in convoluted Hamiltonian  \eqref{hypercycle_d1}.

In the case of $D=1$ hypercycle (cycle graph), $\kappa=k/2+1$ and  Hamiltonian parameters $\gamma_{ij}$ are following: $\gamma_{i,i+1}=\sqrt{2}$ for $i=1$ and $i=k/2$, $\gamma_{i,i+1}=1$ otherwise. This case is depicted in Fig.~\ref{fig:hypercube4}(b), where $5$ waveguides should be used for mapped QW implementation with $k=8$.

\subsubsection{Hypercycle of $D=2$, torus convolution} 
To derive  the formula for $D=2$, we refer to the fact, that hypercycle is obtained as a Cartesian product of circles, so we can write its adjacency matrix in the form:

\begin{equation} \label{hypercycle_cartesian}
    A_{Hypercycle}^{2, k} = A_{Circle}^{k} \otimes A_{Circle}^{k}
\end{equation}

\begin{figure}[ht!]
	\centering
	\includegraphics[width=0.8\linewidth]{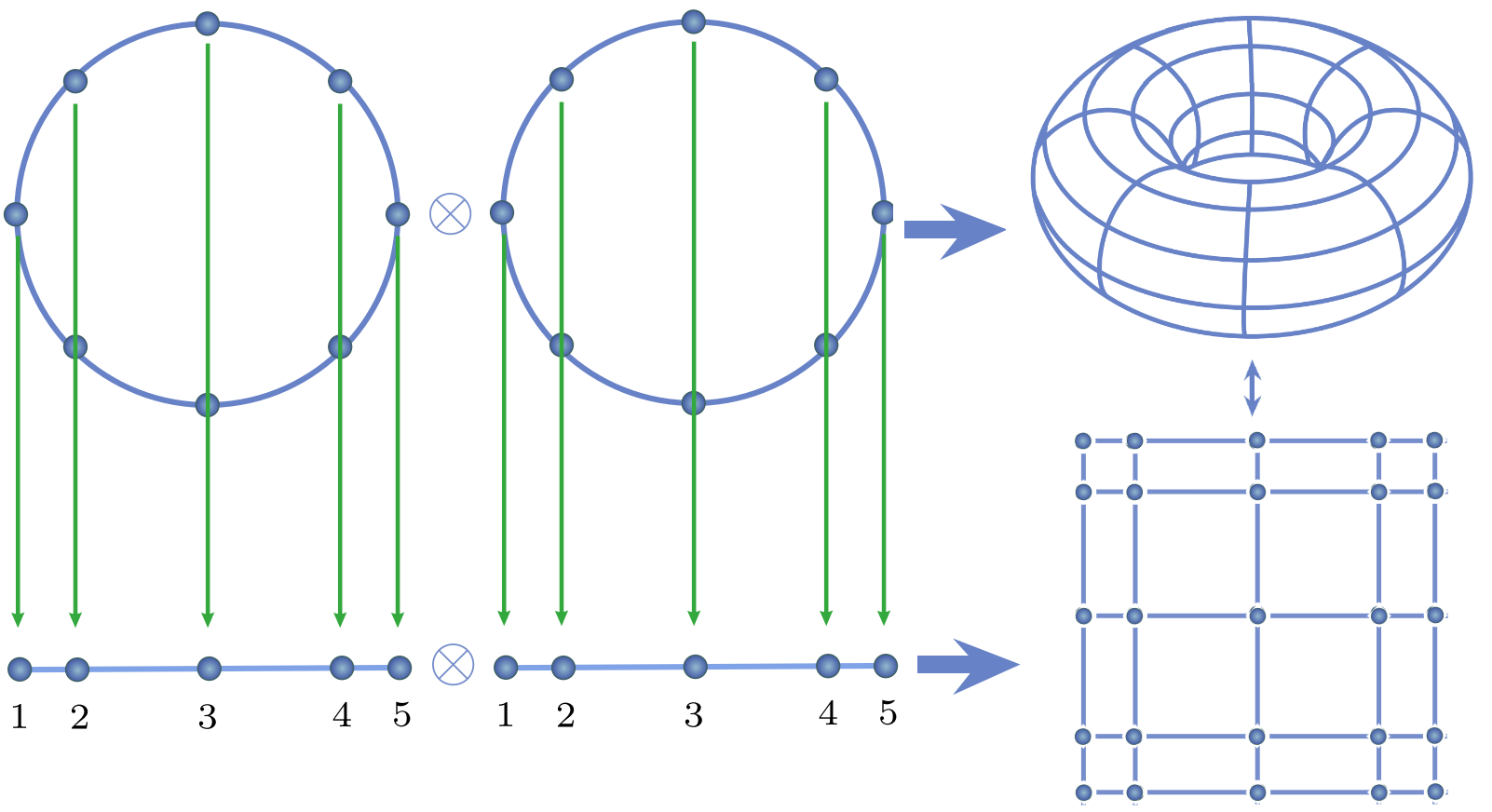}
	\caption{Hypercycle $D=2, k = 8$ (torus) as a Cartesian product of two circles with $k=8$. Transforming them into line graphs and taking a Cartesian product leads to a lattice that is equivalent to the hypercycle in terms of the dynamics of quantum random walk.}
	\label{fig:hypercycle_transform_1}
\end{figure}

We propose that hypercycle in a mapped space could be obtained by mapping individual circles into the lines of  Cartesian product  (Fig. \ref{fig:hypercycle_transform_1}):

\begin{equation} \label{hypercycle_cartesian_mapped}
    A_{Mapped\ Hypercycle}^{2, k} = A_{Mapped\ Circle}^{k} \otimes A_{Mapped\ Circle}^{k}
\end{equation}

To resolve the equation, we utilize the formula for the graph Cartesian product:

\begin{equation} \label{graph_cartesian}
    A(G \times H) = A (G) \otimes I_{|V(H)|} + I_{|V(G)|} \otimes A (H),
\end{equation}
where $|V(H)|, |V(G)|$ are powers of vertices of graphs $H$ and $G$ respectively, $\otimes$ denotes Kronecker product, and $I$ denotes an identity matrix.

Combining  (\ref{hypercycle_d1}) and (\ref{hypercycle_cartesian_mapped}), we obtain at  $D=2$ the following mapping rule for the Hamiltonian:

\begin{equation} \label{hypercycle_d2}
    \begin{aligned}
    \mathcal{H}^\mathrm{hc\rightarrow lattice} = \sum_{i=1}^{\kappa}\gamma_{i,i+1}\big(\ket{i+1}\bra{i}+\ket{i}\bra{i+1}\big)\\ \times \sum_{j=1}^{\kappa}\gamma_{j,j+1}\big(\ket{j+1}\bra{j}+\ket{j}\bra{j+1}\big).
    \end{aligned}
\end{equation}

By using  (\ref{graph_cartesian}), we end up with a formula for mapping QW on hypercycles with $D = 2$:

\begin{equation} \label{hypercycle_d2}
    \begin{aligned}
    \mathcal{H}^\mathrm{hc\rightarrow lattice} = \sum_{i=1}^{\kappa}\gamma_{i,i+1}\big(\ket{i+1}\bra{i}+\ket{i}\bra{i+1}\big) \otimes I_{\kappa}\\ +
    I_{\kappa} \otimes \sum_{j=1}^{\kappa}\gamma_{j,j+1}\big(\ket{j+1}\bra{j}+\ket{j}\bra{j+1}\big),
    \end{aligned}
\end{equation}
which is basically a Cartesian product of two lines. The result of this operation may be established as a ${\kappa}\times{\kappa}$ lattice (Fig. \ref{fig:hypercycle_transform_1}). This lattice is equivalent to a line in the hypercube case;  at the same Hamming distance equiprobable groups would appear. It leads to the same dynamics of QW  on both structures.

\begin{figure}[ht!]
	\centering
	\includegraphics[width=1\linewidth]{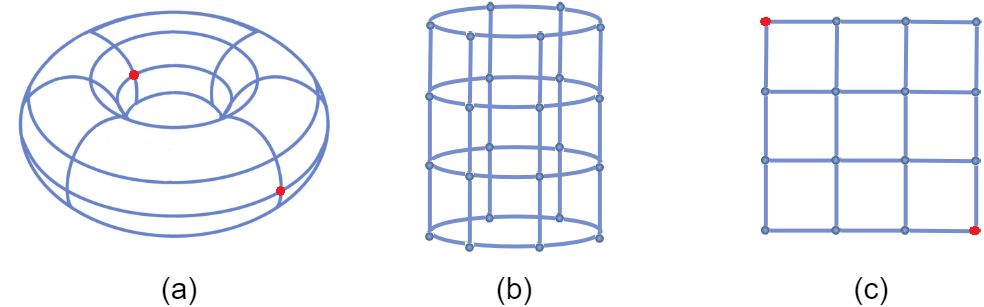}
	\caption{Partial convolution of (a)  torus $D=2$ and $k=6$ into  Kronecker product's figures that leads to different results. In (b) only one of the circles is transformed into the line graph; in (c) both circles are transformed into lines. Red points indicate starting and target nodes used for random walks numerical simulations, performed in Sec. 4.  Sink nodes are not shown in (a) and (b), respectively.}
	\label{fig:hypercycle_equivalence}
\end{figure}

Thus, it is possible to transform only one of the circles in the equation into a line. The overall dynamics on all of the graphs would remain the same. The obtained figure would be a result of the Cartesian product of a line and a circle (Fig. \ref{fig:hypercycle_equivalence}).

\subsection{Lattice convolution}


Consider the problem of reducing a 2D lattice into a smaller graph with fewer nodes and supporting the same dynamical properties as the original one.
As we know,  a square might be transformed into a line graph. It is possible to map a sequence of central squares into the lines reducing the size of the geometric shape. However, this logic does not give us the desired results directly, and the quantum states of a walking particle will differ from one on the original graph (lattice). A proper mapping appears if we remove the lower part of the graph.


\begin{figure}[ht!]
	\centering
	\includegraphics[width=1\linewidth]{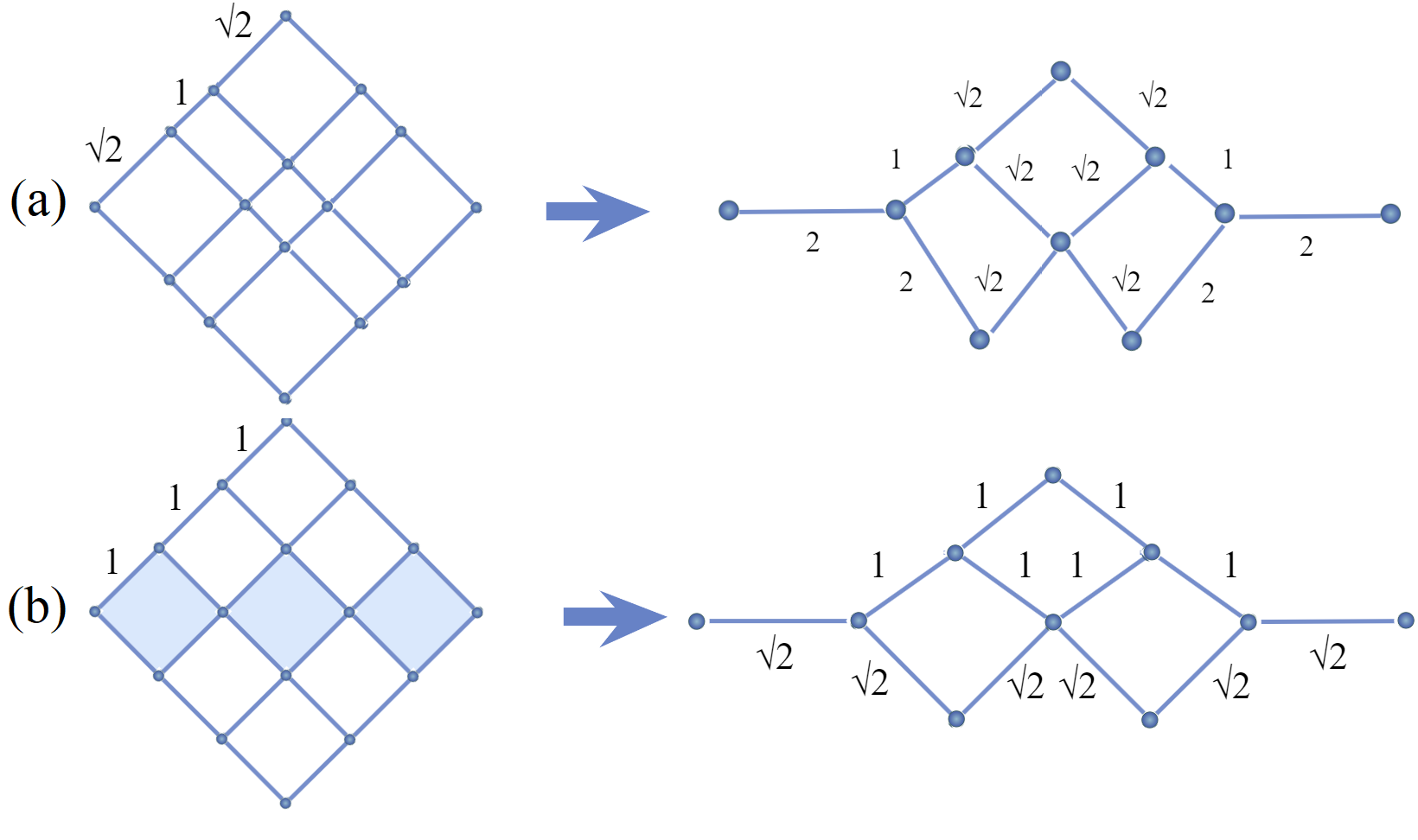}
	\caption{Convolution of $4\times4$ lattices. (a) -- lattice obtained from  hypercycle (torus), see Fig. \ref{fig:hypercycle_transform_1},  and (b) -- arbitrary symmetric 2D lattice, respectively.  Central squares are convoluted and the lower part is removed. The numbers on the edges indicate transition probabilities.} 
	\label{fig:lattice_conv}
\end{figure}

We propose a \textit{heuristic algorithm}  for a lattice convolution that  looks as follows:

\begin{enumerate}
  \item Discard the lower triangular of a lattice;
  \item Convolute diagonal rectangles into lines;
  \item Recompute transitional probabilities between nodes for lines by using the formula: $\sqrt{p_{side1}^{2} + p_{side2}^{2}}$, where $p_{side1}$ and $p_{side2}$ are transition probabilities of two sides of a transformed rectangle.
\end{enumerate}

We have checked this algorithm with different lattices. Fig. \ref{fig:lattice_conv} shows the results of our simulations for two samples of lattices.

In order, Fig. \ref{fig:lattice_conv}(a) demonstrates the convolution of the lattice that represents the convolution of a hypercycle, shown in Fig. \ref{fig:hypercycle_transform_1}.  Central squares of the graph possess edges' probabilities of ${\sqrt2}$ and they transform into 2. 

Thus, the right part of Fig. \ref{fig:lattice_conv}(a) establishes  a smaller  graph  with fewer nodes that may be obtained from the original hypercycle (torus) after two sequential convolutions, which preserve QW dynamical features. 

Notably, the convolution of lattices with higher dimensions is also possible. In  Fig. \ref{fig:lattice_conv}(b) we represent another example that is a convolution of $4\time4$ symmetric lattice into the ultimate graph. 

\section{Numerical experiments}

\subsection{Basic equations}

In this section we implement classical and quantum random walks using a simulation in Python Qutip \cite{johansson2012qutip}. 
We use an approach proposed in  \cite{Melnikov2019}, where an additional sink node for QWs detection is presented. In particular,  we use formulas 
\begin{equation} \label{classical_rw_eq}
    p(t) = e^{(T - I) t}p(0) = e^{-t}e^{Tt}p(0),
\end{equation}  
\begin{equation} \label{quantum_rw_eq}
    \dfrac{d\rho(t)}{dt} = -{i}[\mathcal{H}, \rho(t)] + \Gamma (L\rho(t)L^{\dagger} - \dfrac{1}{2}\{L^{\dagger}L, \rho(t)\}),
\end{equation} 
to examine  dynamics of classical  (Eq.(\ref{classical_rw_eq})) and quantum, (Eq. (\ref{quantum_rw_eq})) random walks,  respectively. In (\ref{classical_rw_eq}), $T$ is a transition matrix, possessing matrix elements $T_{ij}$, which characterize probabilities for a classical particle to jump from $i-$th to $j-$th node.

To examine the Markovian dynamics of QWs we use Gorini–Kossakowski–Sudarshan–Lindblad
(GKSL) master equation  (\ref{quantum_rw_eq}) established for  density matrix $\rho$, e.g. \cite{Manzano}. The Lindblad operator $L$ introduces decoherence in the unitary dynamics described by ${\mathcal H }$, by moving the quantum particle from graph vertex to sink with constant rate. 
In order, the last term in (\ref{quantum_rw_eq}) characterizes decoherence that happens with a quantum particle in Markovian approximation when the particle moves to the sink node.

It is noteworthy, that QWs require an appropriate procedure for detecting the quantum particle. In coherently performed QWs, the quantum particle is "smeared" over all the graph nodes with relevant probabilities. 
To find the particle in the graph's desired  (target) node, we first link the target node with the ancillary sink one as discussed in \cite{Melnikov2019}. 
The particle "falls" into the sink node with a suitable (dimensionless) rate $\Gamma$, indicating the end of the particle walk process on the graph. Large enough $\Gamma\gg1$  implies frequent measurement procedure performed with the particle that mimics quantum Zeno paradox \cite{Gurvitz}. Below we examine the case of  $\Gamma=1$. Physically, this limit means that the rate  $\Gamma$  is compatible with the parameter $\Omega$ that we took equal to 1. 

We assume the sink and its degrees of freedom to be much larger than that of our graph size which justifies neglecting the memory effects in corresponding relaxation processes.  Notably, the examination of various strategies for the detection of QWs (including their possible non-Markovian features) represents a separate problem that we did not attack in this work.

Second, we define hitting detection threshold probability $p_{th}$. In this work we set it at the same level as it was established in  \cite{Melnikov2019}: 
\begin{equation} \label{thresh}
    p_{th} = \frac{1}{\log(n)},
\end{equation}  
where $n$ is a number of vertices of a graph. Thus, if the probability of finding a particle in the target (or, in the quantum case, sink) node is larger than $p_{th}$, it is quite likely that the particle has reached the target.

\subsection{Quantum  and classical random walks dynamics  simulation}

In our first experiment, we aim to determine if there is a variation in the probability of particle detection between the original graphs and their transformations. To achieve this, we numerically examine the dynamics of QWs on both original and transformed samples analysing the hitting probability of the target node for a quantum walker.   The target node is chosen as the furthest node from a starting point based on minimal edge distance; red points in Fig. \ref{fig:hypercycle_equivalence} establish stating and target nodes, respectively.

\begin{figure*}[!t]
	\centering
	\includegraphics[width=1\linewidth]{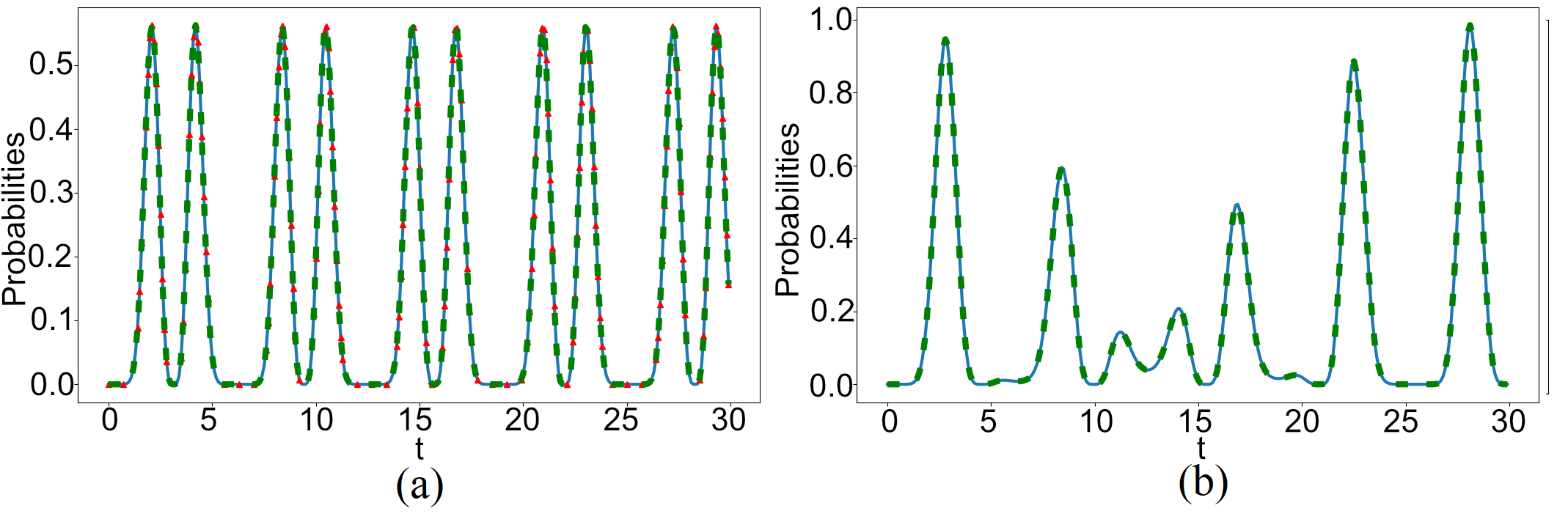}
	\caption{Target node  hitting  probability versus dimensionless time  $t$ for (a) --  hypercycle (torus) with $D = 2$ and $k = 6$ (blue curve), its lattice convolution  (red dot-curve), ultimate graph (green dashed-curve), and for 
  (b)  -- original lattice $4\times4$ (blue curve) and its convoluted (ultimate) graph representation (green dashed-curve).}
	\label{fig:circle_grid_transformed}
\end{figure*}

The obtained results of numerical simulations of QWs performed on hypercycle with $D = 2$, $k = 6$ are established in Fig.  
 \ref{fig:circle_grid_transformed}(a). 
 Three curves correspond to the original torus (Fig. \ref{fig:hypercycle_equivalence}(a)), its lattice convolution (Fig. \ref{fig:hypercycle_equivalence}(c)),  and its transformed (ultimate) graph, cf. Fig. \ref{fig:lattice_conv}(a), respectively. As seen in  Fig. 
 \ref{fig:circle_grid_transformed}(a), all  curves coincide with each other. 

 A similar result is confirmed by  Fig.\ref{fig:circle_grid_transformed}(b),  where we establish simulation of QWs dynamics for $4 \times 4$ (Fig. \ref{fig:lattice_conv}(b)) lattice and its convoluted graph. It is noteworthy, that the target node hitting probability is the same for chosen samples.

 
 Thus, we may conclude, that the proposed method of graph convolution preserves QWs dynamics; it is valid for both lattice and torus cf. Fig. \ref{fig:lattice_conv}. 

The second set of experiments relates to quantum and classical random walks simulation. To characterize quantum and classical random walks on original (torus) and convoluted (lattice) graphs, we compare hitting times for them, measured in a number of steps required to target node arrival. We randomly select pairs of initial and target nodes on the graph, simulate random walks, and compare the time necessary for appearing them at the target node. We assume that a particle possessing random walk occurs at the node if detection probability exceeds the threshold defined in (\ref{thresh}). Only the result of the fastest walk was kept for each simulation; if the walk did not hit the node within a specified period of time, we recognized it as a failure.

\begin{figure*}[!t]
	\centering
	\includegraphics[width=1\linewidth]{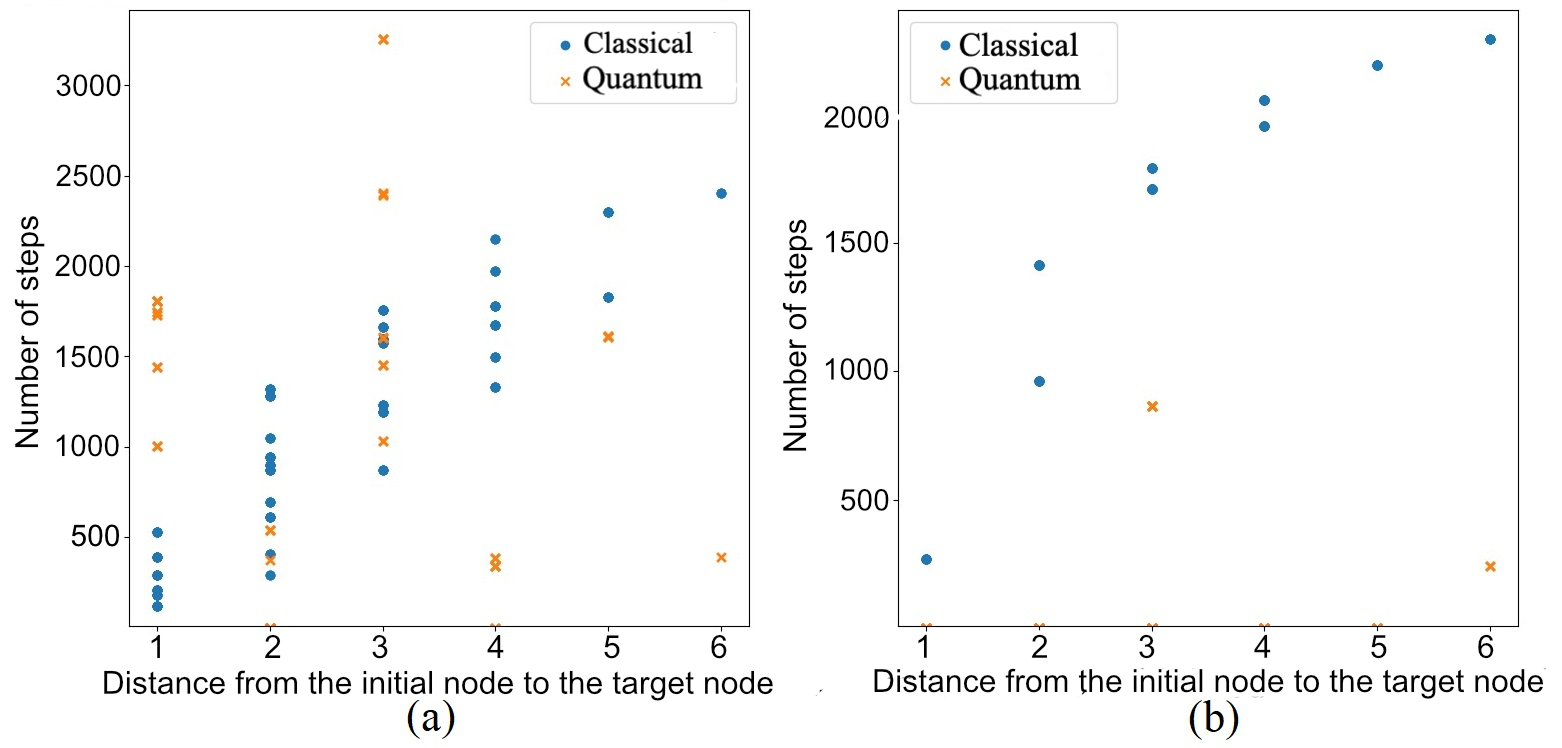}
	\caption{Simulation of quantum-classical random walks on (a) -- convoluted torus $4\times4$ lattice, and (b) -- original torus.   For each run, only the fastest result of the walker is saved, i.e. the one that takes fewer steps to hit the target node.  
 }
	\label{fig:classical_quantum_rw}
\end{figure*}

We performed $300$ simulations on both hypercycle  with $D=2$, $k = 6$ and its $4\times4$ lattice convolution, see Fig.  \ref{fig:hypercycle_equivalence}(a) and Fig. \ref{fig:hypercycle_equivalence}(c), respectively. The results of the numerical experiment for the lattice are depicted in Fig.  \ref{fig:classical_quantum_rw}(a), while the results for the original figure (torus) are represented in Fig.  \ref{fig:classical_quantum_rw}(b).

Notably, due to the different number of nodes the threshold value (\ref{thresh}) differs for original and convoluted graphs. This is why plots in Fig. \ref{fig:classical_quantum_rw}(b) are rare. However, we notice that the points relevant to the original graph can be found in  Fig. \ref{fig:classical_quantum_rw}(a) for its lattice convolution.  

For both the convoluted and original graphs classical and quantum random walks follow the same pattern.
It takes less time for a classical random walk when the distance $d$ is short enough (less than $3$), e.g. \cite{Melnikov2019}. However, for distances $d\geq3$, the quantum walker behaves faster than its classical counterpart.

Generally, the random walk probability for the particle successfully reaching the target node is higher for a lattice structure since the number of nodes is significantly smaller than for the original torus. This property can be used for the improvement of algorithms based on quantum random walks due to the fewer resources exploited.

\subsection{Minimal mapping graph}
For the graphs that we consider in this work,  we determine whether they could be further simplified, or they are already in their smallest configuration.   In general, it depends on the number of equiprobable groups of nodes in the graph. In case when there are fewer groups than the current number of nodes, we can attempt to simplify the structure further. If not, we are not able to reconstruct the QW dynamics of the original sample.

\begin{figure}[ht!]
	\centering
	\includegraphics[width=0.9\linewidth]{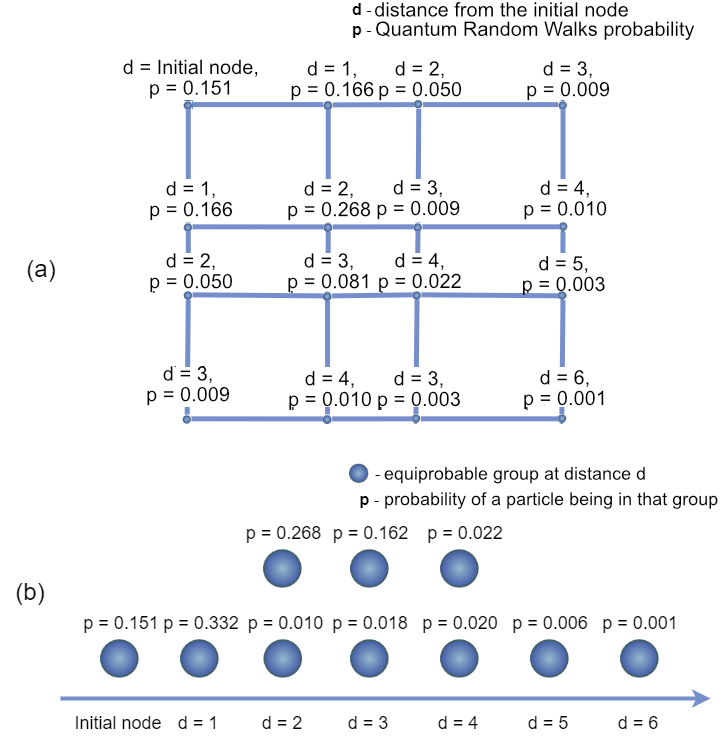}
	\caption{(a) -- probabilities of QW on  $4\times4$ lattice  (convoluted torus) with distances $d$, (b) -- groups of  nodes with equal probabilities.}
	\label{fig:lattice_prob}
\end{figure}

In Fig. \ref{fig:lattice_prob}(a) we examine $4\times4$ lattice that represents the convolution of hypercycle shown in Fig.  
\ref{fig:hypercycle_transform_1}. We represent QW probabilities at the nodes located at distance $d$.  

In  Fig. \ref{fig:lattice_prob}(b) we establish the layout of the equiprobable groups of the nodes for the examined lattice. To determine this layout, we first find the shortest distance from the starting node and the probabilities of QW at each node. Some nodes have the same probability due to the symmetry of the graph. For instance, only two nodes that are located in $d=1$ distance away from the initial node have an equal probability of $0.166$  (see Fig. \ref{fig:lattice_prob}(a)) and are grouped in the left side of  Fig. \ref{fig:lattice_prob}(b); these nodes are unified in the group with probability $2\times0.166=0.332$. For $d=2$, we have two equiprobable groups: two nodes with probability $0.05$ and one node with probability $0.248$. Thus, to preserve the total probability property, probabilities of merging nodes are summed; the sum of all probabilities is equal to one. By repeating this process, we can identify the minimum number of nodes required to represent the original graph, as demonstrated in Fig. \ref{fig:lattice_prob}(b).

The obtained information provides insight into the arrangement of nodes and their number in a convoluted structure for each distance  $d$ but does not reveal the connections between the nodes. Therefore, we can only use this technique for verification, but not for derivation of the convoluted graph.

By comparing Fig. \ref{fig:lattice_prob}(b) and Fig.  \ref{fig:lattice_conv}, we can conclude that the convoluted representation of a lattice is the smallest graph that maintains the dynamical features of QWs of the original lattice if the number of nodes in the convoluted graph is equivalent to the number of equiprobable node groups in the original graph.

\section{Criterion based on adjacency matrices}

Let us switch our attention now to the properties of 
adjacency matrices eigenvalues of the original graph and its convoluted representation. For simplicity we 
restrict ourselves by examination of hypercubes.  For $D = 2$, square, the eigenvalues are 
\begin{equation}\label{original}
    eigenvalues_{original} = [-2, 0, 2, 0].    
\end{equation}
The square convolution, that is three node line graph, possesses eigenvalues  
\begin{equation}\label{mapped}
    eigenvalues_{mapped} = [-2, 0, 2].    
\end{equation}

For the hypercube with $D = 3$  we  have:
\begin{equation}
    eigenvalues_{original} = [3, -3, -1, -1, -1, 1, 1, 1],   \end{equation}
and obtain 
\begin{equation}
    eigenvalues_{mapped} = [-3, -1, 3, 1].    
\end{equation}
for the  four node line  graph that is hypercube $D = 3$ convolution, see Fig. \ref{fig:hypercube3}.

For the hypercycle with $D = 2, k = 8$ (Fig. \ref{fig:hypercycle_transform_1}) we  have:
\begin{equation}
    eigenvalues_{original} = [2, -2, 1.4142, 1.4142, 0, 0, -1.4142, -1.4142],   \end{equation}
and obtain 
\begin{equation}
    eigenvalues_{mapped} = [2, -2, 1.4142, 0, -1.4142].    
\end{equation}


These examples clearly show that the eigenvalues of the convoluted  (line) graph are only unique eigenvalues of the original one.
As a result, we propose (without proving) two observations:

\begin{enumerate}
  \item Equiprobability for the nodes, evidently, occurs due to duplicate eigenvalues in the original matrix.
  \item The minimal number of nodes in the convoluted graph is equal to the number of unique eigenvalues in the original matrix.
\end{enumerate}

It is noteworthy, that we have also seen numerically the validity of these observations for more complex structures, namely the hypercycles considered in the work.

\section{Conclusions}
In this work, we discuss graph convolution problems for lattices, hypercycles, and hypercubes, which presume quantum and classical random walks. We show that it is possible to transform these graphs into smaller ones preserving the QW dynamics. 

We suggest a method for graph convolution based on the fact that some graphs represent a result of Kronecker's product of the line graphs. Theoretically, this approach may be applied to any structure derived through the Kronecker product of two or more graphs. Further, we offer a tool for the convolution of a lattice. We apply the proposed approach for hypercycle (torus) two-step convolution. The first step provides convolution to the lattice. The second one presumes convolution of the lattice to the graph with the minimal accessible number of nodes.   

In this work, we support our approach by using numerical experiments.
The first experiment checks QWs dynamics on original and convoluted graphs. Numerical simulations confirm that the hypercycle (torus),  presuming a two-step convolution procedure, preserves walker quantum dynamics. 
 
In the framework of the second experiment 
we examine hitting times for quantum and classical random walks performed on the original graph (torus) and its lattice convolution, respectively. The dependencies obtained in the work are well consistent with the results obtained earlier, cf. \cite{Melnikov2019}.   

Finally, we propose an approach to obtain the smallest possible graph by using a transformation of the original one. 
We establish a numerically supported link between the eigenvalues and eigenvectors of adjacency matrices, which represent the original graph (hypercube) and its mapping -- the line graph.
We believe that our approach may be explored for graph reduction in more general cases and higher dimensions.

The methods that we discuss in this work can assist in the determination of a minimal number of nodes needed for quantum simulation with photonic waveguides.
The potentially proposed approach allows for the saving of a significant number of qubits required for the QWs simulation due to the symmetry and leads to the creation of more intricate and beneficial algorithms, which are based on the quantum walks paradigm.



\section*{Acknowledgments}
The results of this work related to  Section 2 were funded by the Ministry of Science and Higher Education of the Russian Federation and South Ural State University (Agreement No. 075-15-2022-1116). Studies described in Sections 3 and 4  are supported by project  No. 2019-1339 of the Ministry of  Science and Higher Education of the Russian Federation. 

\section*{Data Availability}

All data used during this study are available within the article.

\section*{Declarations}

\textbf{Conflict of interest:} The authors have no conflicts of interest to declare that are relevant to the content of this article.

\end{document}